\documentclass{article}
\pdfoutput=1

\usepackage{arxiv}

\usepackage[utf8]{inputenc} 
\usepackage[T1]{fontenc}    
\usepackage{hyperref}       
\usepackage{url}            
\usepackage{booktabs}       
\usepackage{amsfonts}       
\usepackage{nicefrac}       
\usepackage{microtype}      
\usepackage{amsmath}
\usepackage{amssymb}
\usepackage{mathtools}
\usepackage{amsthm}

\usepackage{cleveref}       
\usepackage{lipsum}         
\usepackage{graphicx}
\usepackage[round,sort&compress]{natbib}
\usepackage{doi}
\usepackage{subfigure}

\theoremstyle{plain}
\newtheorem{theorem}{Theorem}[section]
\newtheorem{proposition}[theorem]{Proposition}

\theoremstyle{definition}

\theoremstyle{remark}

\renewcommand{\headeright}{}
\renewcommand{\undertitle}{}
\renewcommand{\shorttitle}{Response Time in Human Choice Prediction}

\title{Response Time Improves Choice Prediction and Function Estimation for Gaussian Process Models of Perception and Preferences}

\author{ Michael Shvartsman\\
Meta Reality Labs Research \\
\href{mailto:michael.shvartsman@meta.com}{michael.shvartsman@meta.com}
	\And
Benjamin Letham\\
Meta \\
\href{mailto:bletham@meta.com}{bletham@meta.com}
    \And
Stephen Keeley\\
Department of Natural Science\\
 Fordham University \\
\href{mailto:skeeley1@fordham.edu}{skeeley1@fordham.edu}
}

\begin{document}
\maketitle 

\begin{abstract}
Models for human choice prediction in preference learning and psychophysics often consider only binary response data, requiring many samples to accurately learn preferences or perceptual detection thresholds. The response time (RT) to make each choice captures additional information about the decision process, however existing models incorporating RTs for choice prediction do so in fully parametric settings or over discrete stimulus sets. This is in part because the de-facto standard model for choice RTs, the diffusion decision model (DDM), does not admit tractable, differentiable inference. The DDM thus cannot be easily integrated with flexible models for continuous, multivariate function approximation, particularly Gaussian process (GP) models. We propose a novel differentiable approximation to the DDM likelihood using a family of known, skewed three-parameter distributions. We then use this new likelihood to incorporate RTs into GP models for binary choices. Our RT-choice GPs enable both better latent value estimation and held-out choice prediction relative to baselines, which we demonstrate on three real-world multivariate datasets covering both human psychophysics and preference learning applications.
\end{abstract}

\section{Introduction}

Human binary choice data is widely used to measure latent mental constructs. Key motivating applications are human psychophysics, the study of human perception in psychology \citep{Kingdom2016a}; human value-based decision making \citep{rangel2008framework}; and preference learning \citep{furnkranz03}. In all cases, humans give binary responses about whether they detect a stimulus or can discriminate between two stimuli (in psychophysics), or about which of two options they prefer (in value-based decision making and preference learning). While binary choice experiments have been used in psychology for more than a century \citep[e.g.][]{fechner1860}, they have seen many recent advances in the machine learning community, particularly through nonparametric latent function modeling and active learning. Since the work of \citet{chu05}, Gaussian processes (GPs) have been a standard approach in preference learning for modeling latent utility functions from binary expressed preferences defined over general multivariate and continuous feature spaces. Among their many applications, human preference data have been used to learn robot locomotion policies \citep{tucker21, polar,  cosner22}, personalize assistive devices \citep{thatte17, tucker20}, and learn a good golf swing \citep{biyik20}. Recent work in machine learning for psychophysics has similarly used GP models to learn latent perceptual functions from binary human feedback, for purposes including evaluating hearing loss \citep{gardner2015bayesian, gardner2015psychophysical} and understanding visual sensitivity to image distortion \citep{guan22, letham2022look}.

In all of these applications, the model assumes that binary responses derive from a latent function of stimulus configuration that is mapped to choice probability through a sigmoidal link function. There are two important aspects of the problem that are detrimental to the sample efficiency of the model. First, information is lost for large portions of the latent space that are mapped to choice probabilities very near 1 or 0. Second, areas of the function with high uncertainty, i.e.\ regions where preference or detection probability is near 0.5, require many samples for accurate estimation as they have large Bernoulli variance, $p(1-p)$. These shortcomings are largely due to the fact that binary responses alone are a very coarse reflection of the subject's decision process. A richer model of the decision process should allow for discrimination between a `yes' response with choice probability close to 0.5 and one close to 1. 

Mathematical psychology and computational neuroscience provide rich models for the underlying decision process. These models incorporate additional information, notably response times, as a way of inferring the underlying latent variables leading to the subject's response \citep[e.g.][]{Clithero2018a,Laming1968a, Donders1969a, Sternberg1969a}. One of the most popular such models is the diffusion decision model\footnote{Equivalently, the drift-diffusion model.} \citep[DDM;][]{Bogacz2006a,Ratcliff1978a,ratcliff2008diffusion}. Unfortunately, the joint choice-RT likelihood under the DDM cannot be computed in closed form. A variety of numerical approaches can be used to approximate it \citep[e.g.][]{navarro2009fast,Voss2008}, but none are differentiable. They thus cannot be incorporated into a modern variational GP approximation framework in a straightforward way. Our core contribution is to approximate the DDM using a family of parametric skewed distributions, which enables for the first time the use of variational GP models with DDM-inspired RT-choice likelihoods. 

We study the performance of the model using both simulated data from synthetic problems and data from real human subject studies. On synthetic problems, we show that incorporating RTs into the model provides more accurate estimation of the latent function than choice-only models, especially in the low-data regime. On real data, we show that incorporating RTs into GP models for human perception and preferences can substantially improve choice prediction performance relative to choice-only models. This is the case even when the choice probability is the only quantity of direct interest and the RTs are solely used as side information for the modeling, as in ML applications in this domain.

Section \ref{sec:background} provides background on RT modeling and the DDM. Section \ref{sec:gp} introduces the GP classification model used for modeling human choices. Section \ref{sec:rt_choice} then describes our novel DDM approximation and how we use that to jointly model RTs and choices in a GP. Section \ref{sec:synthetic_exp} describes the synthetic experiments, followed by the real-world psychophysics and preference learning experiments in Sections \ref{sec:csf} and \ref{sec:preference} respectively. Our psychophysical dataset is a high-dimensional visual psychophysics task taken from  \citet{letham2022look}. Our first preference learning example is a study of recommender system evaluation, containing pairwise evaluations of A/B test outcomes at a large internet company \citep{bope}. Our second preference learning example is a novel robotics preference learning dataset, where a participant evaluated pairs of simulated robot gaits to identify locomotion parameters that produced the most natural looking gait. 

\section{Background}\label{sec:background}

\begin{figure}
\includegraphics[width=\textwidth]{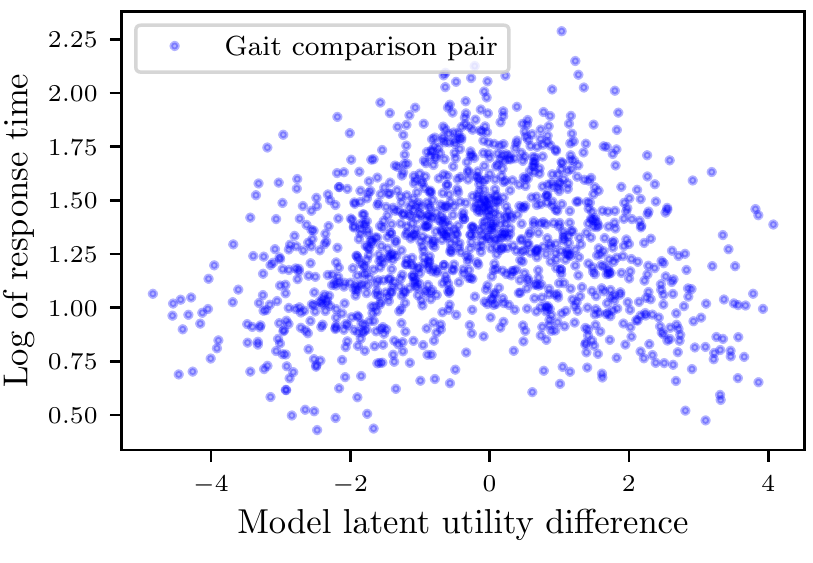}
\caption{A human subject evaluated 1,225 pairs of robot gaits to select the more natural looking gait. The response time in making the judgement was longer for pairs that have a small difference in their modeled latent utilities, reflecting the increased challenge of judging between gaits of similar quality. Response time can serve as useful side information for predicting the probability of a subject's choice.}
\label{fig:rt_vs_f}
\end{figure}

As noted above, RTs are well-studied in psychology and computational neuroscience \citep[e.g.][]{Laming1968a, Donders1969a, Sternberg1969a}, and are empirically known to correlate with choice probability. To illustrate this, Fig.~\ref{fig:rt_vs_f} shows RT data from the multivariate robot gait optimization task we study in Section~\ref{sec:robot}, in which a human subject watched two simulations of a quadruped robot walking, each with different gait parameters, and was asked which gait looked more natural. The figure compares the latent GP utility estimates for each evaluated pair using the binary preference data only (`choice-only' model), with the response time of the human subject in judging that pair. When the difference in latent utility for the pair is 0, they are equally preferred and the choice probability is 0.5. Gaits with closer utility values had both longer and more variable response times, while those with large differences in utility were easier to judge and had shorter and less variable response times. This is the relationship that we use to improve latent function estimation and choice prediction.

Our starting point for modeling RTs is the DDM. While many other models exist, the DDM is widely used for modeling decision making in neuroscience and psychology, and can be motivated from a variety of theoretical perspectives: as a generalization of classical signal detection theory in psychophysics \citep{Ashby1983a,Griffith2021a}, as a sequential statistical inference process \citep{Bogacz2006a}, as an approximation to neural firing rates \citep{Gold2002a,Gold2007a}, or as a mechanistic theory of memory \citep{Ratcliff1978a}. With just a few parameters, the model describes the joint distribution of choices and reaction times. The response time is generally understood to reflect a process of evidence accumulation, sequential statistical inference, or integration over neural noise. When this process reaches some threshold determined by the desired accuracy of the decision maker, a decision is made. The process completes more quickly when stronger signal is available, resulting in faster decisions when signal is stronger.

Our objective is to combine the flexibility of GP models for perception and preferences with the domain knowledge about RTs encoded in the DDM. To do so, we propose a model (Section~\ref{sec:rt_choice}) that places a GP prior on the drift parameter of the DDM. Larger latent values (stronger preferences, or more clearly perceived stimuli) correspond to shorter and less variable RTs. Latent GP values map to choice probabilities via a sigmoid link whose specific functional form we discuss below. 

Existing DDM models almost universally estimate parameters independently over a set of discrete experimental conditions, making them incompatible with the general continuous stimulus spaces that we consider here. This is in large part because the likelihood of the RT distribution under the DDM is the first passage time of a 1-d Wiener process with nonzero drift and nonzero initial condition to one of two boundaries. While expressions for this density are well known \citep{feller1967introduction}, they take the form of an infinite summation. This sum can be truncated to control approximation error depending on specific parameter values \citep{navarro2009fast} but naive application of this approximation is incompatible with modern automatic differentiation frameworks: first, because bounding density error does not necessarily bound error in gradients; and second, because varying term counts per parameter value make it difficult to leverage standard batched linear algebra operations. Alternate approaches solve the Kolmogorov backward equation associated with the DDM diffusion process \citep{Voss2008,Shinn2020,voss2015assessing} or approximate parameters by moment-matching to the data \citep{van2017ez}. Given this complexity, standard approaches to DDM estimation rely on full MCMC using slice sampling \citep[e.g.][]{frank2015fmri} or gradient-free optimization. 

Our focus is not necessarily improving DDM likeihood approximation or density estimation. Rather, we would prefer to use a simpler density that is still able to represent the latent value or signal strength we need for choice prediction in a GP framework. Unfortunately, while other distributions have been used to describe response times, their parameters do not straightforwardly map to the domain knowledge encoded in the DDM process \citep{matzke2009psychological}. Instead of using such distributions directly, we use the fact that closed-form expressions for the conditional moments of the DDM distribution are known even if the exact density is intractable \citep{srivastava2016explicit}. We use these moments, which are a function of the DDM parameters, to match the the moments of a shifted, skewed distribution with a known functional form such as the shifted lognormal or shifted inverse gamma distributions. We select the parameters of these three-parameter distributions to uniquely match the mean, variance, and skew of the DDM distribution \citep{lo2014momenta}. 

\section{GP Models for Human Choices}\label{sec:gp}
GP-based models can successfully model psychophysical tuning curves \citep{gardner2015bayesian, owen2021adaptive,letham2022look,KeeleySemiP}, as well as latent preference values \citep{chu05,bope}. Observations are given as $\{\mathbf{x}_n,y_n\}_{n=1}^N$, where $\mathbf{x}_n \in \mathbb{R}^d$ are the parameters of the multi-dimensional stimulus configurations, and $y_n \in \{0, 1\}$ are subject responses. The typical GP approach to modeling in this setting is to assume a latent function $z$ with a GP prior:
\begin{equation}\label{eq:z_gp}
z(\mathbf{x}) \sim \mathcal{G P}\left(0, k_\theta \left(\cdot,\cdot \right)\right).
\end{equation}
For single choices (e.g.\ `stimulus present' vs `stimulus absent'), the kernel $k_\theta(\mathbf{x},\mathbf{x'})$ can be a standard GP kernel such as the radial basis function (RBF), which we use throughout our experiments. For choices between paired stimuli (e.g.\ `prefer 1' or `prefer 2'), we assume that the choice probability is determined by the difference between the values assigned to the two stimuli \citep{chu05}, which means that the prior over value differences remains a GP with a `preference kernel' given by \citet{houlsby2011a}. In both cases we estimate hyperparameters controlling the amplitude, and an independent lengthscale per input dimension (i.e., an ARD kernel), $\theta_{G} = \{\rho, \mathbf{\ell}\}$. 

The observation model is Bernoulli, and assumes that $y$ is conditionally independent of $\mathbf{x}$, given $z$. Formally, let $z_n = z(\mathbf{x}_n)$, and $y_n \sim \textrm{Bernoulli}(\Phi(z_n))$, where $\Phi(\cdot)$ is the Gaussian cumulative distribution function. Prior work has varied the choice of the sigmoid and the details of the kernel, but has maintained this basic model structure. We are primarily interested in inferring $z$, both for the purpose of predicting $y$ and for extracting useful information such as detection thresholds and most-preferred configurations. In the experiments in this paper, we refer to this model as the `choice-only' model since it uses only choice data $y_n$. We now show how this model can be extended to incorporate RT observations.

\section{The RT-choice Model}\label{sec:rt_choice}
We augment the GP model above to include a distribution over response times. Here, our data are $\mathcal{D} = \{\mathbf{x}_n,y_n, t_n\}_{n=1}^N$ where $\mathbf{x}_n$ and $y_n$ are as before, and $t_n \in [0, \infty )$ are subject response times. As before, we assume the corresponding latent function values $z_n$ are a function of the stimulus configuration $\mathbf{x_n}$, and put a GP prior on $z(\mathbf{x})$ as in (\ref{eq:z_gp}).

\begin{figure*}[ht!]
\centering
\includegraphics[width = .9\textwidth]{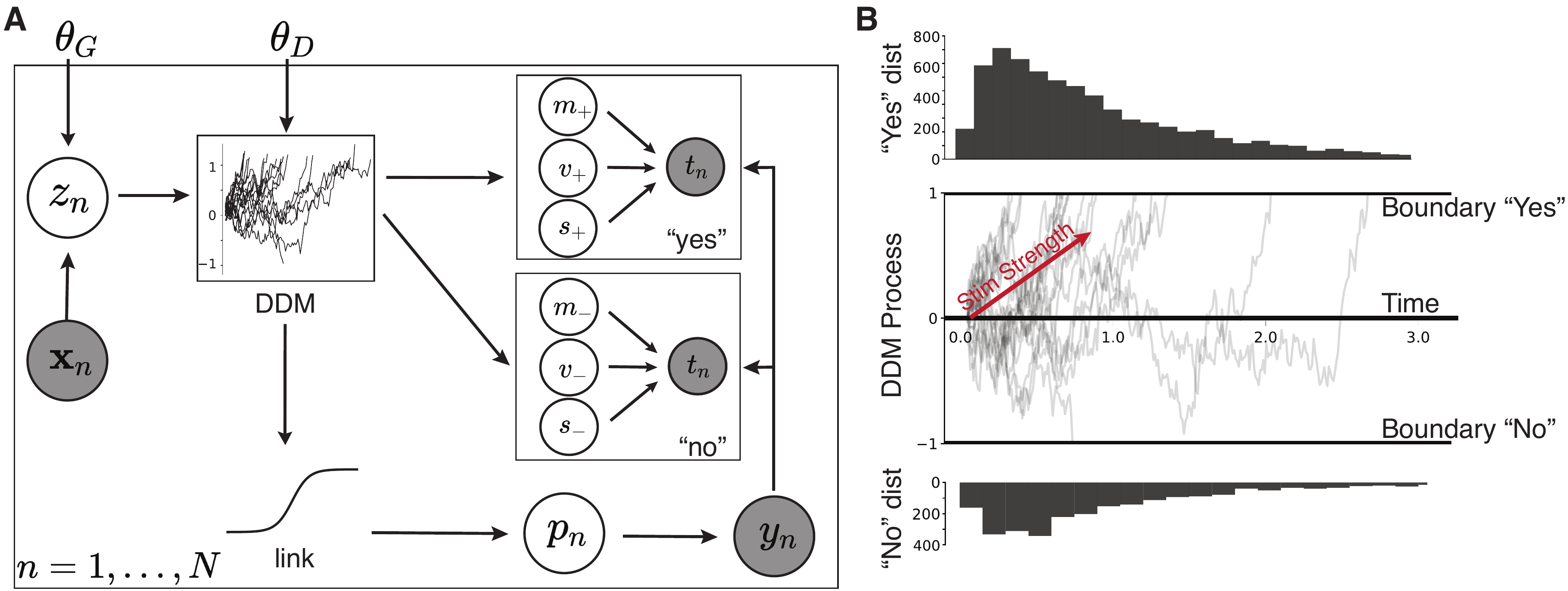}
\caption{\textbf{A.} Graphical depiction of our RT-choice model. The latent function, characterized with a GP prior, is given as the drift parameter in the DDM and mapped to both reaction time distribution moments as well as a choice probability at every location in a high-dimensional stimulus space.  \textbf{B.} Schematic of the drift diffusion model. A stochastic process with an average drift (red arrow) dictates random movement in a latent space, capturing an underlying binary decision making process. The latent accumulator eventually reaches one of two boundaries, representing one of two binary decisions, providing both a choice and a response time. RT histograms for each choice (top and bottom) are skewed distributions with known empirical moments. }
\label{fig:graphical}
\end{figure*}

Let $\mathbf{y}=(y_1, y_2, \dots, y_N) $, $\mathbf{t}=(t_1, t_2, \dots, t_N) $, $\mathbf{z}=(z_1, z_2, \dots, z_N)$, and $\mathbf{X}=(\mathbf{x}_1, \mathbf{x}_2, \dots, \mathbf{x}_N)$. The joint likelihood of response times and choice responses can be written as:
\begin{align}\nonumber
p(\mathbf{t},& \mathbf{y}\mid  \mathbf{X}, \theta_{G}, \theta_{D})\\\label{eq:likelihood} & =\int p(\mathbf{t}\mid \mathbf{z}, \mathbf{y}, \theta_{D}) p(\mathbf{y} \mid \mathbf{z}, \theta_{D})p(\mathbf{z}|\mathbf{X},\theta_{G})d\mathbf{z}.
\end{align}
Here both response times $\mathbf{t}$ and choices $\mathbf{y}$ are assumed to depend on the stimuli $\mathbf{X}$ only via the latent function $\mathbf{z}$; see Fig. \ref{fig:graphical}A for a graphical representation of the model. The distribution of the latent $p(\mathbf{z}|\mathbf{X},\theta_{G})$ will be Gaussian due to the GP prior on $z$. The choice distribution, $p(\mathbf{y} \mid \mathbf{z}, \theta_{D})$, and the conditional RT distribution, $p(\mathbf{t}\mid \mathbf{z}, \mathbf{y}, \theta_{D})$, we specify according to a DDM, which has parameters $\theta_{D}$. We will now describe the DDM and these distributions.

\subsection{The Drift Diffusion Model}\label{sec:ddm}
The DDM can be simulated as a Wiener process that stochastically moves towards one of two boundaries, the `yes' boundary and the `no' boundary. Whichever boundary is reached first is the choice made, $y_n$, and the time required to reach the boundary is the RT, $t_n$. An illustration of the DDM process is shown in Fig. \ref{fig:graphical}B. The movement of the process towards a boundary models the accumulation of evidence, and when the boundary is reached, there is sufficient evidence to make a a judgement. The RT is thus the first passage time of this process, a well-studied quantity in stochastic processes.

The DDM contains several parameters: drift rate, the decision threshold level ($C$), the particle initial condition ($x_0$), and a shift ($t_0$). We use the GP latent function value $z_n$ as the drift strength, providing an explicit link between the stimulus configuration $\mathbf{x}_n$ and the response produced by the DDM. The remaining parameters, $\theta_{D} = \{C, x_0, t_0\}$, will be directly estimated from data.

The DDM process induces different RT distributions for the `yes' and the `no' choices, depending particularly on the sign and strength of the drift parameter as it favors one choice over the other. Evaluating the likelihood under these distributions is intractable, but their moments can be computed analytically as a function of $\theta_{D}$ and $z_n$ \citep{srivastava2016explicit}. The first three moments of the RT distributions are denoted in Fig. \ref{fig:graphical} as ($m_+$, $v_+$, $s_+$) and ($m_-$, $v_-$, $s_-$) for the mean, variance, and skew of the 'yes' and 'no' distributions, respectively. We now describe how the DDM distributions are incorporated into the model marginal likelihood in (\ref{eq:likelihood}).

\subsection{The Choice Distribution}
\begin{figure}
\includegraphics[width = .9\columnwidth]{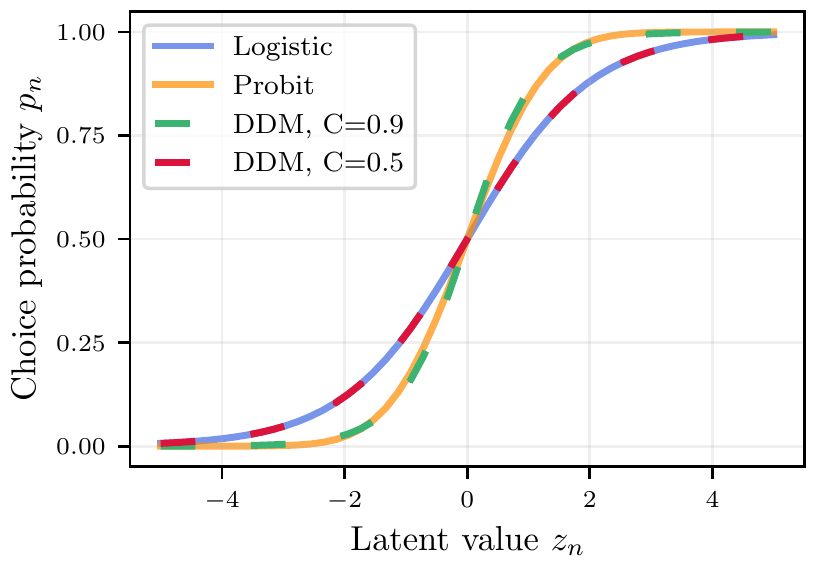}
\vspace{-10pt}
\caption{The DDM link function closely matches typical probit and logistic link functions, depending on the process parameters.}
\label{fig:links}
\end{figure}

The choice distribution in (\ref{eq:likelihood}), $p(\mathbf{y}|\mathbf{z}, \theta_D)$, is the distribution of first boundary crossings from the DDM process. We assume conditional independence across trials,
\begin{equation*}
    p(\mathbf{y}|\mathbf{z}, \theta_D) = \prod_{n=1}^N p(y_n|z_n, \theta_D),
\end{equation*}
and have $y_n | z_n, \theta_D \sim \textrm{Bernoulli}(p_n)$. The DDM process induces the following link function between the latent function values and the choice probability \citep{srivastava2016explicit}:
\begin{equation}\label{eq:ddm_choice}
    p_n = \frac{e^{2 C z_n} - e^{-2 x_0 z_n}}{e^{2 C z_n} - e^{-2 C z_n}}.
\end{equation}
Fig. \ref{fig:links} shows how this link function compares to the probit and logistic sigmoid link functions that have been used in choice-only GP models, for $x_0 = 0$. It can closely match either depending on the DDM boundary parameter $C$.

\subsection{Moment Matching the RT Distribution}\label{sec:ddm_moments}
The conditional RT distributions as a function of choice, $p(\mathbf{t}\mid \mathbf{z}, \mathbf{y}, \theta_{D})$, are not available in closed form under the DDM process, however, as discussed above, the moments are. To obtain a tractable likelihood in (\ref{eq:likelihood}) we will assume conditional independence across trials,
\begin{equation*}
    p(\mathbf{t}\mid \mathbf{z}, \mathbf{y}, \theta_{D}) = \prod_{n=1}^N p(t_n|z_n,y_n, \theta_D),
\end{equation*}
and will then use a parametric distribution for $p(t_n|z_n,y_n, \theta_D)$, whose parameters are set by moment matching to the DDM RT distribution.

\begin{proposition}[\protect{\citealt{srivastava2016explicit}}]\label{prop:moments}
Let $k_{z} = C z_n$ and $\tilde{y}_n = k_{z} + x_0 z_n (-1)^{(1 - y_n)}$. The RT distribution under the DDM process has as its moments:
\begin{align*}
     \mathbb{E}[t_n | &z_n, y_n, \theta_D]\\ 
    &= t_0  +  \frac{1}{z_n^2}\Big(2 k_z \operatorname{coth}\left(2 k_z\right)- \tilde{y}_n \operatorname{coth}\left(3 k_z-\tilde{y}_n\right)\Big),
\end{align*}
\vskip -0.2in
\begin{align*}
   \textrm{Var}[ &t_n | z_n, y_n, \theta_D] = \frac{1}{z_n^4}\Big(4 k_z^2 \operatorname{csch}^2\left(2 k_z\right)\\ &+ 2 k_z \operatorname{coth}\left(2 k_z\right) 
-\tilde{y}_n^2 \operatorname{csch}^2\left(\tilde{y}_n\right) -\tilde{y}_n \operatorname{coth}\left(\tilde{y}_n\right)\Big),
\end{align*}
\vskip -0.2in
\begin{align*}
    \textrm{Skew}[&t_n | z_n, y_n, \theta_D] = \frac{1}{z_n^6} \Big(12 k_z^2 \operatorname{csch}^2\left(2 k_z\right)\\  &+16 k_z^3 \operatorname{coth}\left(2 k_z\right) \operatorname{csch}^2\left(2 k_z\right) +6 k_z \operatorname{coth}\left(2 k_z\right)\\ &-3\tilde{y}_n^2 \operatorname{csch}^2\left(\tilde{y}_n\right)  -2\tilde{y}_n^3 \operatorname{coth}\left(\tilde{y}_n\right) \operatorname{csch}^2\left(\tilde{y}_n\right) \\&-3\tilde{y}_n \operatorname{coth}\left(\tilde{y}_n\right)\Big).
\end{align*}
\end{proposition}
We use these three moments from the DDM to match to parametric RT distributions. Considering that RT distributions are typically heavy-tailed \citep{murata2014stochastic}, we focus here on heavily skewed distributions for our parametric RT forms. In our experiments, we use the lognormal, shifted lognormal, shifted inverse gamma, and shifted gamma distributions. In the economics community, expressions are available for the parameters of these distributions as a function of the empirical sample statistics---specifically the mean, variance, and skew \citep{lo2014momenta, brignone2021moment}. These expressions allow for analytic moment matching with the known expressions of the DDM RT moments above. To evaluate the likelihood, we use the current DDM parameters and GP function samples to compute the mean, variance, and skew of the RT distribution according to Prop. \ref{prop:moments}. The parameters of the desired parametric RT distribution are then computed from those moments via moment matching, and we evaluate the likelihood of the RTs under this parametric distribution. The formulae for computing the parameters from the moments for each skewed distribution are provided in the supplementary materials. This moment-matched parametric distribution is then used as the RT component of the likelihood in (\ref{eq:likelihood}), $p(t_n|z_n,y_n, \theta_D)$.

\begin{figure}
\begin{center}
\centerline{\includegraphics[width=\columnwidth]{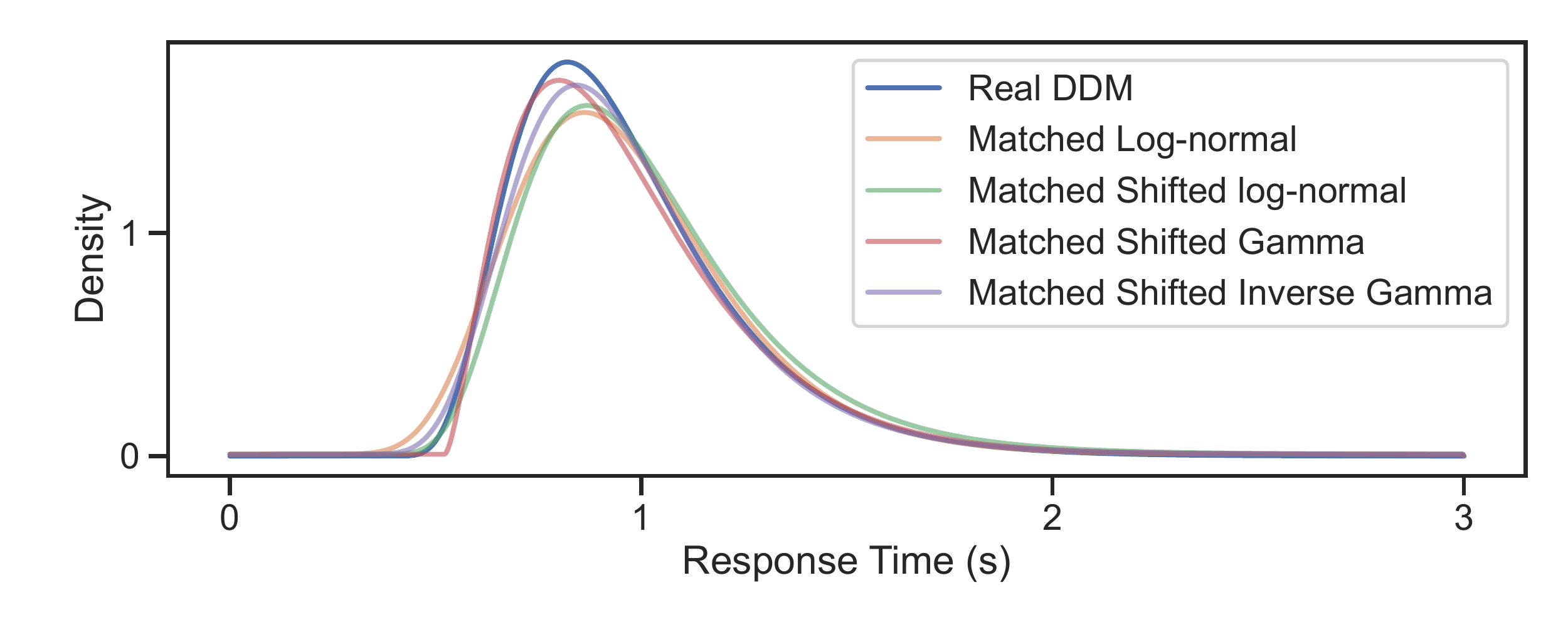}}
\caption{Example of the real DDM distribution (computed using the truncated infinite series approximation of \protect{\citealt{navarro2009fast}}), and our moment-matched approximations. Moment-matching can closely approximate the true DDM RT distribution.}
\label{fig:momentmatch}
\end{center}
\end{figure}

Figure \ref{fig:momentmatch} shows how the numerically calculated DDM RT distribution is captured by each of our parameterized heavy-tailed distributions via moment-matching Prop.~\ref{prop:moments} with the parameter expressions given in the supplement. While subtle differences in densities are apparent on visual inspection, the overall shapes are similar, and we will see below that these approximations are of sufficient quality to enable RT-choice to outperform the choice-only model. For additional discussion of these parameterizations, see the supplement.

\subsection{Inference}
Because the marginal likelihood in (\ref{eq:likelihood}) cannot be computed in closed form, we use standard variational methods to approximate the GP posterior, and obtain point estimates of $\{\theta^{G}, \theta^{D}\}$). Since the distribution over our latent function is Gaussian, we can use Gauss-hermite quadrature in the expectation term of the traditional evidence lower bound, and optimize the objective using standard methods \citep{Hensman2015bb, balandat2020botorch}.

\section{Synthetic Experiment}\label{sec:synthetic_exp}
\begin{figure}[ht!]
\begin{center}
\centerline{\includegraphics[width=.9\columnwidth]{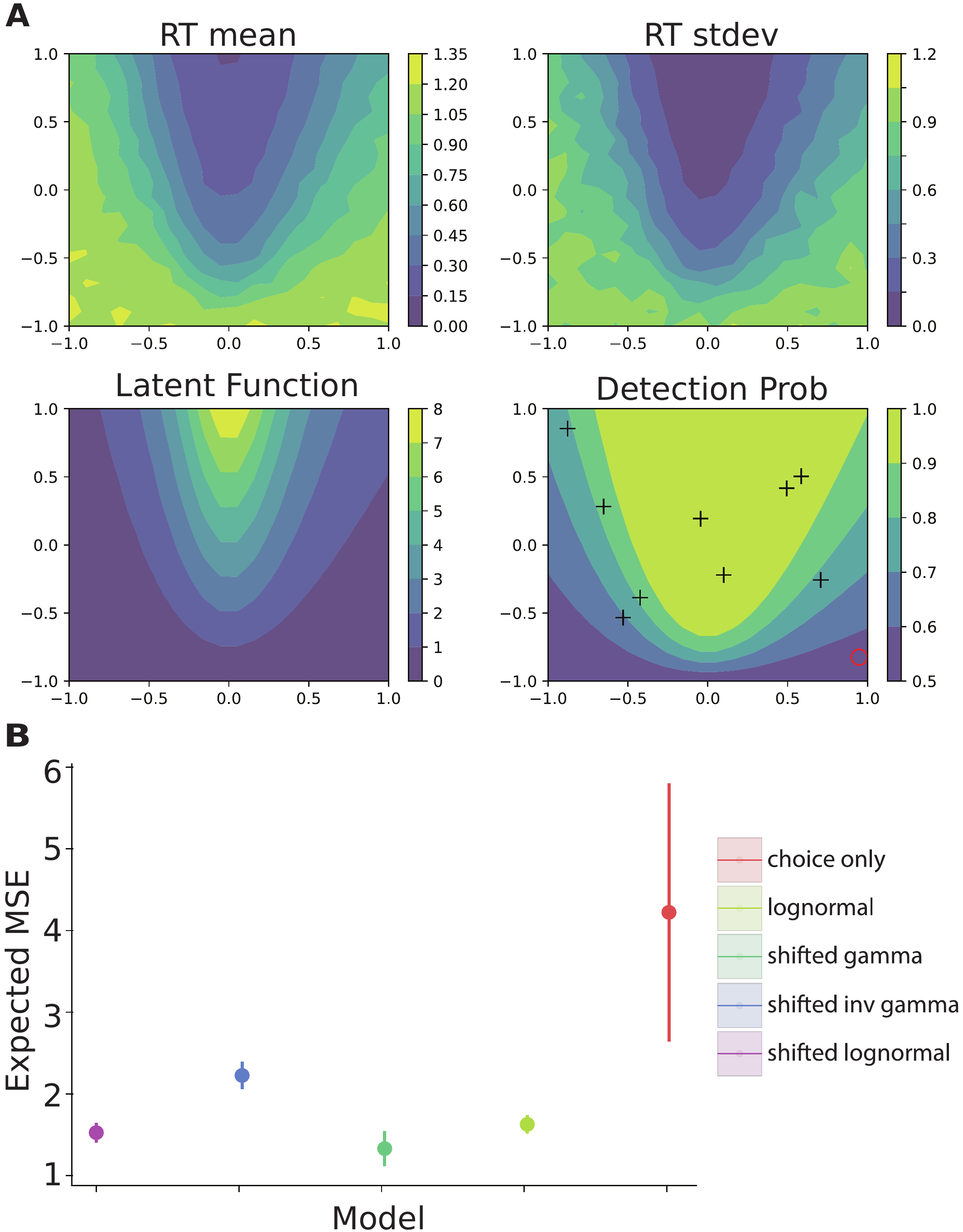}}
\caption{\textbf{A.} The mean and standard deviation (top) of the response time in our 2D test function, as well as its latent function value (bottom left) and associated choice probability (bottom right). From ten samples taken at random positions, we compare latent recovery using the RT-choice and choice-only models. \textbf{B.} Mean squared error in expectation over the posterior of the latent function under each model fit on 10 observations. Error bars show standard error over 10 simulated datasets.}
\label{fig:hairtie_synth}
\end{center}
\end{figure}

To demonstrate the benefits of our approach, we begin with a synthetic data experiment. We use the 2-d detection test function of \citet{owen2021adaptive}, which was designed to evaluate models for psychophysics, scaled by a factor of 0.2.  We performed the scaling because typical drift rates in the literature are often found in that range \citep{matzke2009psychological} and RTs with very large drifts are difficult to accurately simulate numerically as they yield near-instantaneous responses. Fig~\ref{fig:hairtie_synth}A shows the basic properties of the test function. At every point in the parameter space, the test function (bottom left) was used as the drift parameter of a DDM. From this underlying latent function and the DDM parameters, we can calculate the mean and standard deviation of the RTs for each stimulus using Prop. \ref{prop:moments} (top row). The latent function values also generate choice probabilites via (\ref{eq:ddm_choice}) (bottom right). Shown atop the bottom right panel in Fig. \ref{fig:hairtie_synth}A are example stimulus locations $\mathbf{x}_n$ in our 2-d parameter space. At each, the choice $y_n$ and the RT $t_n$ were obtained by simulation of the DDM process as described in Sec.~\ref{sec:ddm}. The simulated choice $y_n$ is shown for these stimuli in the figure, indicated with a $``+"$ for a correct detection and $``\circ"$ for detection failure. 

Note that there is only a single negative response (detection failure) in this example, a common occurrence in the low-data regime in such problems. In this case, a choice-only model cannot do much more than separate the space into broad `yes' and `no' regions, whereas a model taking advantage of response times can perform far better. To illustrate this point, Fig.~\ref{fig:hairtie_synth}B shows error on recovering this true function from only 10 observations, in expectation over the GP posterior. All variants of the response time model far outperform the choice-only model in this regime, and the choice model's performance is highly variable as it strongly depends on the presence of sufficiently balanced numbers of `yes' and `no' trials. As we will see below, this advantage for the RT models in the low-data regime persists in real-world problems. 


\section{Real-World Psychophysics}\label{sec:csf}

As a first evaluation of our model in a real-world setting, we fit the model to response times and choices in a high-dimensional visual psychophysical task. These data consist of 1,500 trials from a two-alternative forced choice (2AFC) task \citep{letham2022look}, and we obtained response times by contacting the authors. For each trial, the participant was presented with an animated circular Gabor patch, one half of which had been scrambled. The scrambled side was randomly selected on each trial, and the subject was asked to select which side contained the non-scrambled stimulus. The stimulus in each trial varied along six dimensions (contrast, background luminance, temporal and spatial frequency, size, and eccentricity), rendering some trials harder than others in a high-dimensional space. The purpose of the study was to determine how visual perception depends on those six stimulus properties, and to extract detection thresholds from the latent function. Additional dataset details and an example stimulus are available in the supplement. 

\begin{figure*}[ht!]
\centering
\includegraphics[width = \textwidth]{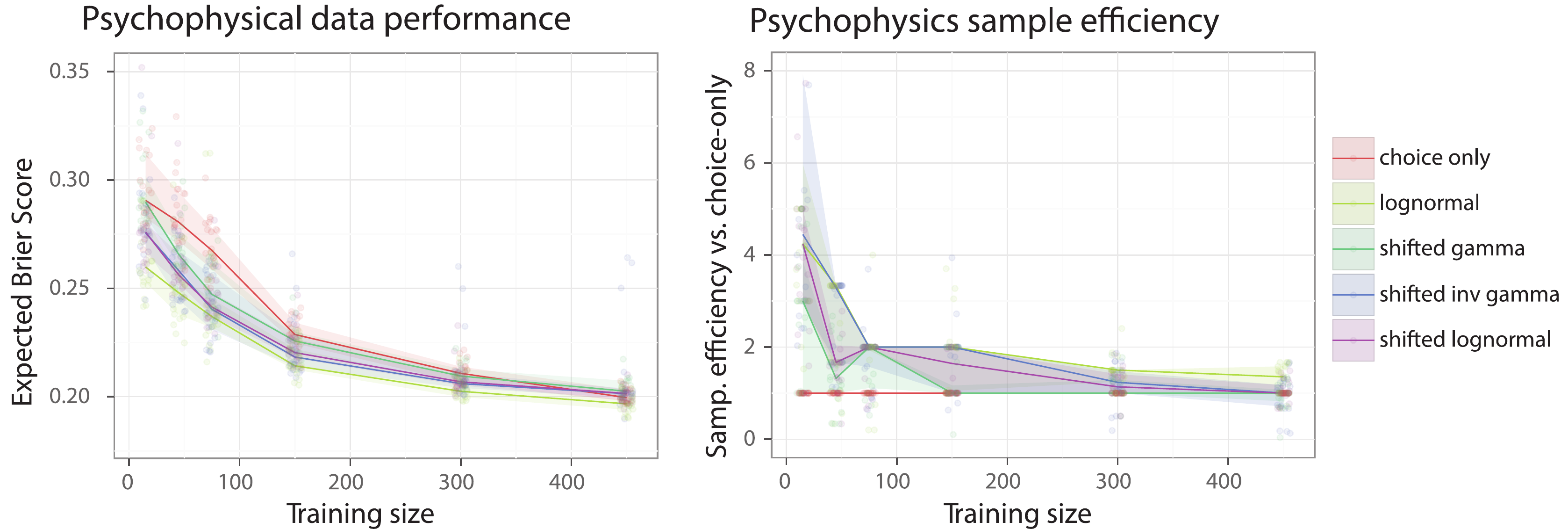}
\caption{\textit{Left} Cross-validated expected Brier score on visual psychophysics data. \textit{Right} Sample efficiency factor: the amount of data needed in the choice-only model to achieve equal performance to the RT-choice models. Lines are medians, shaded areas are IQRs, and individual scores are jittered slightly for better visibility.}
\label{fig:Mikes_data_full}
\end{figure*}

We study how model performance varies with the amount of data.
For each evaluation, we selected a small subset of trials for training (up to 450), and tested on the remainder, shuffling for 10 random folds of cross-validation. We evaluated the same set of models as in the synthetic problem: 4 variants of RT-choice, and the choice-only model. Fig.~\ref{fig:Mikes_data_full} shows expected Brier score of the held-out data. The Brier score \citep{brier1950verification} is a proper scoring rule, and evaluating it in expectation over the model's posterior measures the calibration quality of the model's predictions. We see that our three-parameter RT-choice models consistently outperform the choice-only model, especially for small numbers of trials ($N<200$). 

To more precisely evaluate the sample efficiency gains of the RT-choice models, we additionally computed a `sample efficiency multiplier' by finding the number of samples at which the choice-only model reached the same score as a given RT-choice model (by linear interpolation). The right side of Fig.~\ref{fig:Mikes_data_full} shows this multiplier: the choice-only model is a solid line at 1x, and the RT-choice models achieve multipliers as large as 4x for small data, meaning that they performed as well as the choice-only model did with 4 times less data. 

\section{Real-World Preference Learning}\label{sec:preference}
We next evaluate our model on pairwise data for preference learning, in which instead of the subject selecting an option based on underlying latent perceptual strength, selection is based on which option has higher value with respect to a latent preference or utility function. This means that the underlying latent function value $z$ is queried only via pairwise comparisons. Previous work has shown that this is equivalent to the setting we have already considered, using a different GP kernel specifically designed for pairwise comparisons \citep{houlsby2011a}. Thus, to use our model for preference learning problems, we simply replace the ARD RBF kernel in the GP prior with the preference kernel of \citet{houlsby2011a}. 
\begin{figure}
\begin{center}
\centerline{\includegraphics[width=.9\columnwidth]{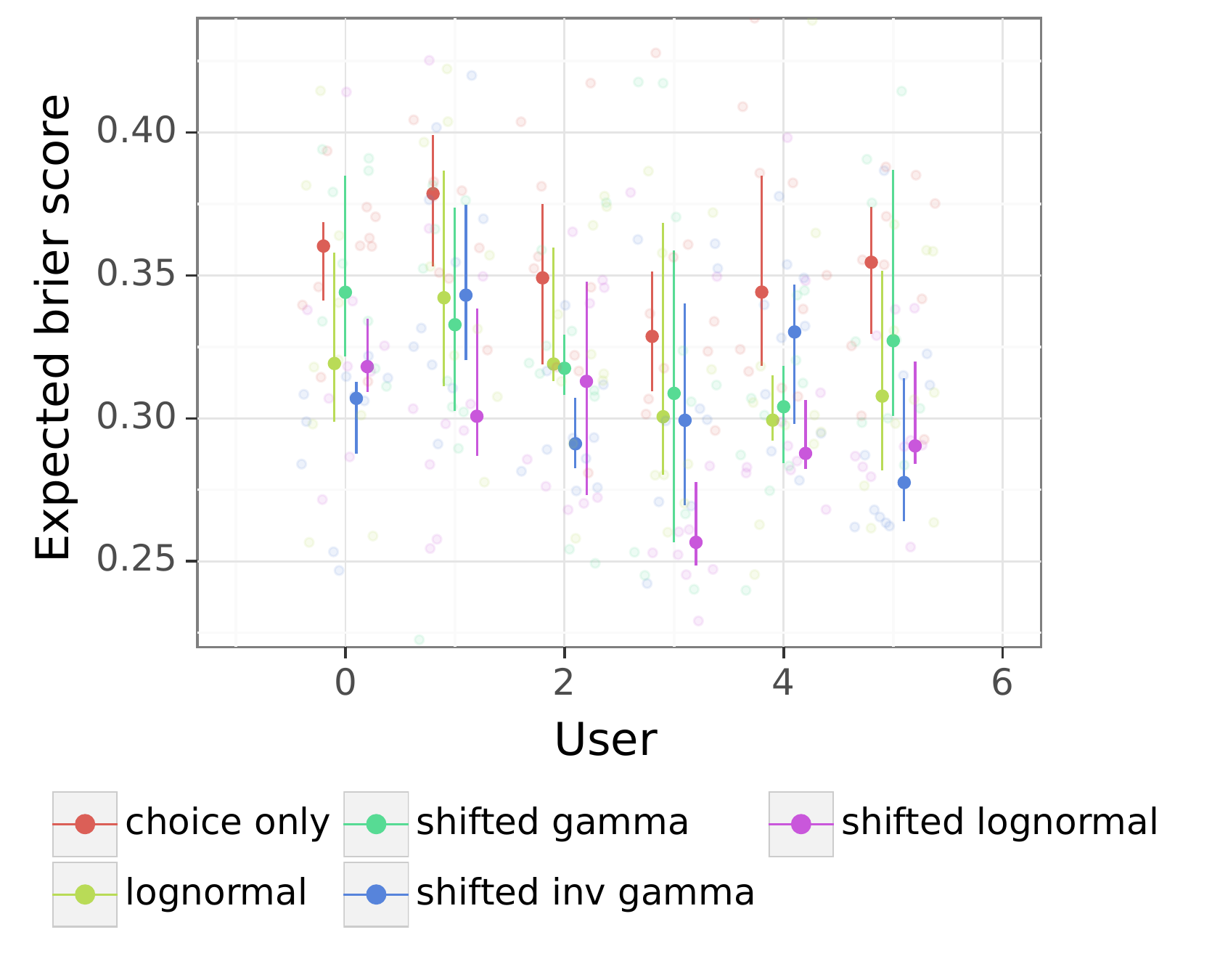}}
\caption{Cross-validated Brier score on preference learning for recommender system evaluation. Points are medians, lines are IQRs, and individual scores are jittered slightly for better visibility.}
\label{fig:Jerry}
\end{center}
\vskip -0.2in
\end{figure}

\begin{figure*}[ht!]
\centering
\includegraphics[width = \textwidth]{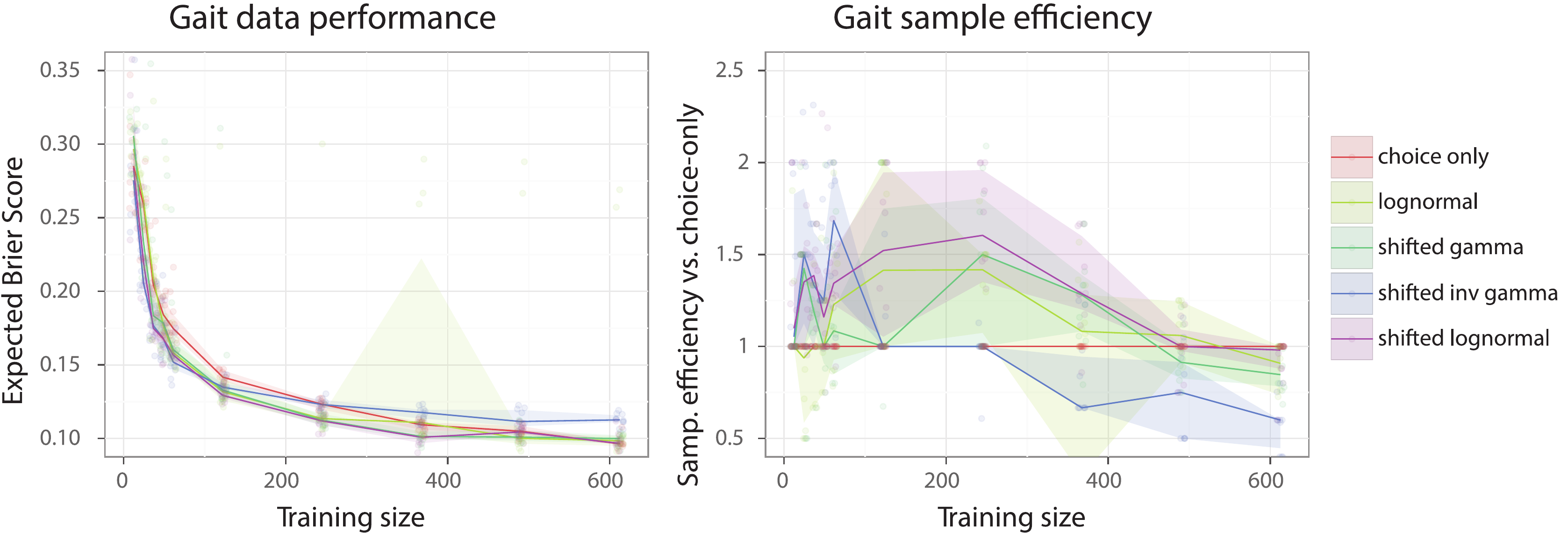}
\caption{\textit{Left} Cross validated expected Brier score on robot gait preference data. \textit{Right} Sample efficiency multiplier for robot gait preference data, relative to choice-only model. Lines are medians, shaded areas are IQRs, and individual scores are jittered for better visibility. }
\label{fig:robot_full}
\end{figure*}
\subsection{Preference Learning for Recommender System Evaluation}\label{sec:bope}
The data for this task come from the user study of \citet{bope}, in which six data scientists and machine learning engineers at a large internet company were asked to compare pairs of A/B test results that showed performance of a recommender system under different configurations. For each pair, the subject was asked to identify which test had the better outcome, for the purpose of finding the most-preferred recommender system configuration. This was a challenging task because the results for each A/B test included changes in up to 9 metrics related to the performance of the recommender system, and the subject had to weigh the relative benefits of changes in these various metrics. We obtained response time data by contacting the authors. Additional dataset information is in the supplement. 

We generated 10 random splits of the data, training on 80$\%$ and testing on 20$\%$ to compute expected Brier score on the held-out data. The number of trials for each subject in this study was comparatively small ($N<50$), so we used all trials per subject in each split of the data. Fig~\ref{fig:Jerry} shows that RT-choice models outperformed the choice-only model for all 6 subjects. The best parametric form for the RT distribution varied across subjects, between the shifted inverse gamma, lognormal, and shifted gamma distributions. However, for all subjects, all of the RT-choice models outperformed the choice-only model.

\subsection{Preference Learning for Robot Gait Optimization}\label{sec:robot}
This problem explored a 3-d space of gait parameters for a simulated Spot quadruped robot \citep{spotminimini2020github}. The parameters were the swing period, step velocity, and clearance height. Using OpenAI Gym \citep{gym}, 10 second videos were recorded of gait simulations for each of 50 quasirandom points in the gait parameter space. A single human participant consented to data collection, and evaluated each of the 1,225 possible pairings of videos to identify which gait appeared more natural. Videos were shown simultaneously side-by-side, so the subject could respond whenever they had determined which was the more natural gait. Response times were recorded for each evaluation, as shown in Fig.~\ref{fig:rt_vs_f}. See the supplementary material for more details about the parameter space and a screenshot of the experiment UI. The goal of the experiment was to produce the most natural-looking gait for the robot, according to the human subject.

As with the psychophysical dataset, we estimated the model from small fractions of the data (up to 600 trials) and evaluated it on the remainder. As with the other preference learning dataset, we used the preference kernel of \citet{houlsby2011a}. Remaining fitting and evaluation details were identical to the psychophysical dataset. Fig.~\ref{fig:robot_full} demonstrates that as in the other datasets, models taking advantage of response times outperformed the choice-only model in the low-data regime, though as in the psychophysical dataset this advantage disappeared by the time about 300 trials had been collected. The sample efficiency gains (Fig.~\ref{fig:robot_full}, right) are somewhat more modest, but the response time models still achieve comparable performance to the choice-only model in as little as half the data. 

\section{Discussion}
We have demonstrated that models that take into account the response-time distribution improve latent function estimation and held-out predictive accuracy in both psychophysics and preference-learning settings. By using the moments of the RT distributions provided in closed form by the DDM, we can calculate point estimates of parameters of a parametric density over RTs, and leverage this additional information to better predict human performance and understand latent cognitive representations. Of the four parametric densities that we evaluated here, the best choice varied across experiments. In practice, cross-validation can be used to select the best approximating density for a particular problem, which is feasible due to model fitting requiring only a few seconds for the training set sizes used here.

We show results in both a synthetic setting and a broad variety of real-world scenarios: human visual psychophysics, preferences in recommender system evaluation, and robot gait tuning. We note improvement specifically in the regime where the number of samples per-subject is small ($N<100$ samples). Improvements in this small-sample regime have practical utility, as it can be time-consuming and uncomfortable for humans to participate in experiments for hundreds or thousands of trials. 

Our work is limited, however, for this very same reason. We show no cross-subject prediction and no pooling or utilizing of data across multiple subjects \citep{Wiecki2013a}. Combining cross-participant pooling, flexible GP models, and use of RT distributions may enable future practitioners to even better predict binary human choices in few samples.

\textbf{Ethics Statement} Our work carries low-risk of ethical harm, as it focuses on binary responses in simple decision making tasks in low-sensitivity settings.  For this work, we only consider de-identified data where subjects have provided explicit informed consent, and we keep our conclusions narrow and pertinent to model performance. We draw no broad conclusions about general human behavior. We anticipate minimal risk associated with future application of our work.

\textbf{Computational load} All benchmarks we report required $<50$ hours on a single Amazon EC2 HPC node. Single model fits, which are most relevant for future practitioners, take on the order of seconds on a typical laptop. 

\pagebreak

\bibliographystyle{unsrtnat} 
\bibliography{Bibliography,mike}

\newpage
\appendix
\onecolumn

\setcounter{section}{0}
\setcounter{equation}{0}
\setcounter{figure}{0}
\setcounter{table}{0}
\setcounter{page}{1}
\renewcommand{\thesection}{S\arabic{section}}
\renewcommand{\thetable}{S\arabic{table}}
\renewcommand{\thefigure}{S\arabic{figure}}

\section{Response Time Information Improves Latent Value Estimation and Prediction of Human Choices: Supplementary Materials}

\subsection{Inference details}
\begin{figure}[htb]
\begin{center}
\centerline{\includegraphics[width=\textwidth]{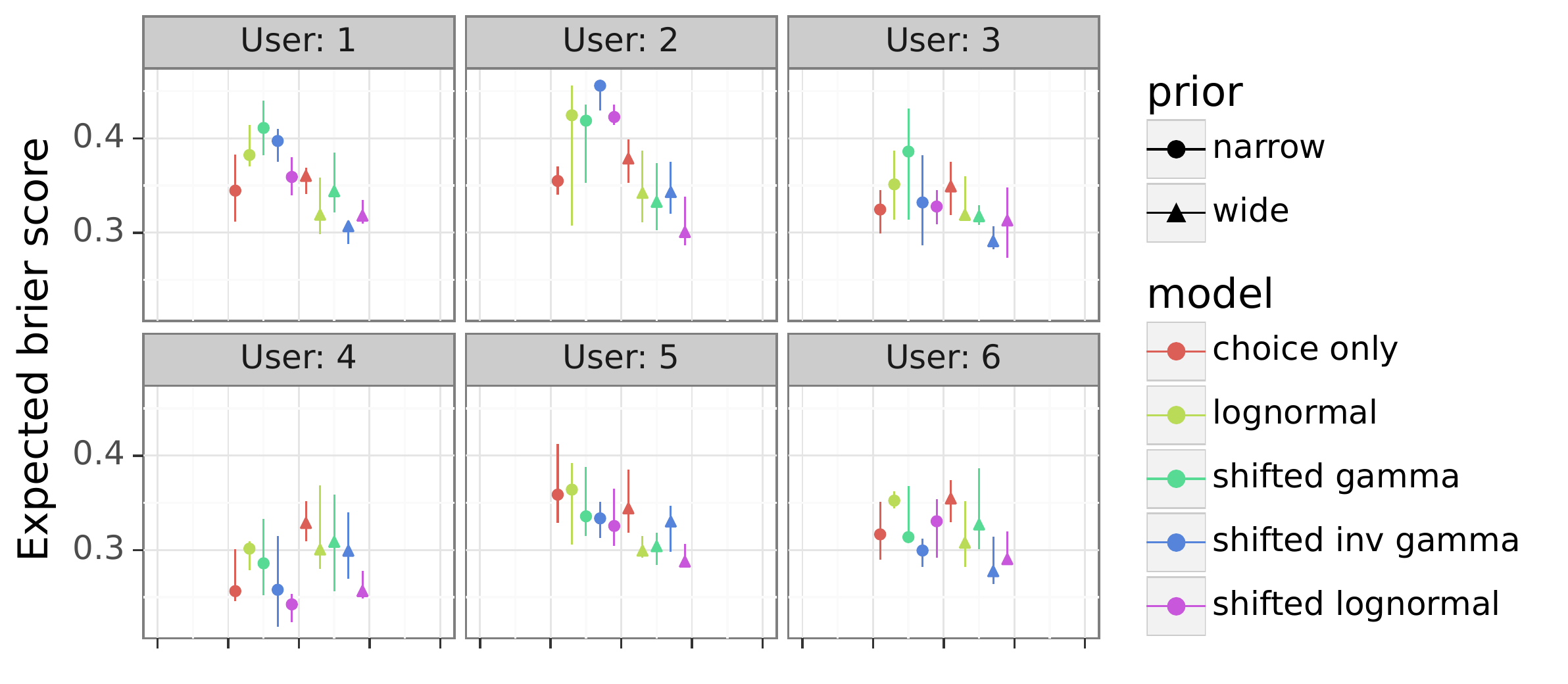}}
\caption{Recommender system performance study, comparing wide to narrow priors.}
\label{fig:jerry_priors}
\end{center}
\end{figure}

As noted in the main text, we used standard variational methods for approximate GPs \citep{Hensman2015bb, balandat2020botorch}. In all cases we used the Adam optimizer \citep{Kingma2014a} with a learning rate of 0.01 and 5000 iterations. Inputs were normalized to $[0, 1]$. The hyperprior for lengthscale was InverseGamma($4.6, 1.0$), selected because it restricts approximately 95\% of the prior probability mass to be between 0.1 and 0.5 (i.e.\ excluding very short or long lengthscales relative to the normalized input domain). The Hyperprior for the variance was selected as Uniform(1, 4), as it restricts the GP output to values that are not saturated by the probit sigmoid, and we wanted to keep priors consistent between the models for consistency. However, we found that the restricted variance prior provided for relatively poorer fits to the recommender system dataset for the RT model while not substantially improving choice model performance (see Fig.~\ref{fig:jerry_priors}). We suspect this is because in that dataset, response times are much longer than in the remaining datasets, choices are less noisy, and a wider range of drift rates are needed. Therefore, we replaced this prior with a much wider Uniform($1$, $10^5$) (essentially unconstrained) prior only for that dataset. Importantly, while the narrow prior hurts the RT-choice models more than it hurts the choice-only model in this setting, if we select the best-performing prior for each model, the RT models still outperform the choice-only model (i.e.\ it does not change our overall conclusion). We report the wider-prior results for simplicity in the main text. 

We report medians and IQRs instead of means and standard errors in the main text because models sometimes failed to converge, creating outliers (which we still show as raw data in the main text). 

\subsection{Dataset details}

\subsubsection{Human psychophysics dataset}
\begin{figure}[htb]
\begin{center}
\centerline{\includegraphics[width=0.3\textwidth]{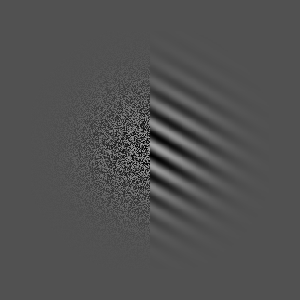}}
\caption{Example psychophysics stimulus (reprinted from \protect{\citealt{letham2022look}})}
\label{fig:csf_example}
\end{center}
\end{figure}

This dataset is as reported in \citet{letham2022look}, except response times are included. It consists of 1500 observations of a single participant making detection judgments. Stimulus features were contrast, pedestal (background luminance), temporal frequency, spatial frequency, size, and eccentricity. An example stimulus is shown in Fig.~\ref{fig:csf_example}. Response times ranged from 0.16 to 15.87 seconds.

\subsubsection{Recommender systems dataset}
This dataset is as previously reported in \citet{bope}, except with response times included. The dataset consisted of data from seven participants whose response times ranged from 4 to 429 seconds. Table~\ref{tab:jerry-data} includes additional information about this dataset. Data from the participant with only 20 observations were not used, since the test set size would be 4 instances only. 

\begin{table}[htb]
\centering
\caption{Dataset details for recommender systems dataset.}
\begin{tabular}{lll}
Participant ID & Instances & Dimensions  \\
1 & 20 & 7 \\
2 & 41 & 8 \\
3 & 41 & 11\\
4 &43 & 5\\
5 &50 & 6\\
6 &50 &  7\\
7 & 50 & 8\\
\end{tabular}
\label{tab:jerry-data}
\end{table}

\subsubsection{Robot gait preference learning dataset}
The simulation framework was from \citet{spotminimini2020github}, and we built on the demo simulation from the package. The selected parameters and their ranges were taken from the package demo settings. Specifically, SwingPeriod ranged from 0.1 to 0.4; StepVelocity from 0.001 to 3; and ClearanceHeight from 0 to 0.1. All other gait parameters and all settings related to the simulation itself were fixed to defaults.

\begin{figure}
\begin{center}
\centerline{\includegraphics[width=\columnwidth]{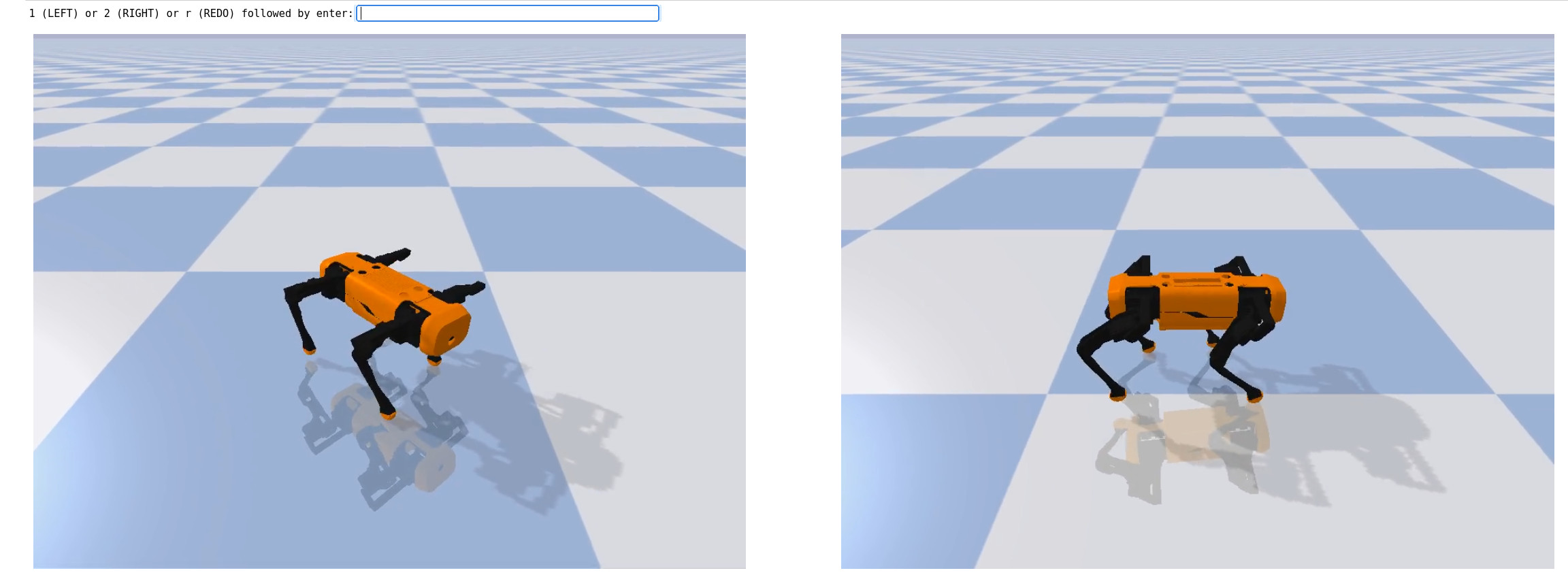}}
\caption{A screenshot of the UI for the robot gait preference learning experiment of Section \ref{sec:robot}. The subject viewed two videos simultaneously playing side-by-side, and selected the one with the more natural gait. Both choice and response time were recorded to fit models of gait preference.}\label{fig:robot_ui}
\end{center}
\end{figure}

A total of 50 simulation videos were recorded for the study, and all pairs were compared by one rater. Fig. \ref{fig:robot_ui} shows a screenshot of the UI for the study in which the human participant viewed two gaits side-by-side and selected the more natural looking. Response times ranged from 1.54 to 9.85 seconds.

\subsection{Moment match info}

Here we provide the probability density functions of all heavy-tailed reaction time distributions we use in this work. Of these, the lognormal is the only two-parameter distribution whereas the shifted gamma, shifted inverse gamma and the shifted lognormal are all three-parameter distributions. The sample statistics calculated from the reaction times, specifically the mean, variance, and skew, are denoted $m_*$, $v_*$, and $s_*$, respectively. Expressions for the three-parameter distributions below are adapted from \citet{lo2014momenta}.
\\

\subsubsection{Shifted lognormal}
$$
f(z ; \mu, \sigma, \eta)=\frac{1}{\sigma(z-\eta) \sqrt{2 \pi}} \exp \left\{-\frac{(\ln (z-\eta)-\mu)^2}{2 \sigma^2}\right\}, \quad z>\eta
$$
with with parameter estimates as a function of sample statistics

$$
\hat{\mu}=\ln \left(m_*-\eta\right)-\frac{\sigma^2}{2}, \quad \hat{\sigma^2}=\ln \left|1+\frac{v_*}{\left(m_*-\eta\right)^2}\right|, \quad \hat{\eta}=m_*-\frac{\sqrt{v_*}}{s_*}\left[1+(B)^{\frac{1}{3}}+(B)^{-\frac{1}{3}}\right]
$$
$$
B \equiv \frac{1}{2}\left(s_*^2+2-\sqrt{s_*^4+4 s_*^2}\right) \in(0,1]
$$
\subsubsection{Shifted inverse gamma}
$$
f(z ; \alpha, \beta, \eta)=\frac{\beta^\alpha}{\Gamma(\alpha)}\left(\frac{1}{z-\eta}\right)^{\alpha-1} \exp \left\{-\frac{\beta}{z-\eta}\right\}, \quad z>\eta, \beta>0
$$
with parameter estimates as a function of sample statistics
$$
\begin{aligned}
\hat{\eta} &=m_*-\frac{\sqrt{v_*}}{s_*}\left[2+\sqrt{4+s_*^2}\right] \\
\hat{\alpha} &=2+\frac{\left(m_*-\eta\right)^2}{v_*} \\
\hat{\beta} &=\left(m_*-\eta\right)(\alpha-1)
\end{aligned}
$$
\subsubsection{Shifted gamma}
$$
f(z ; \alpha, \beta, \eta)=\frac{(z-\eta)^{\alpha-1}}{\beta^\alpha \Gamma(\alpha)} \exp \left\{-\frac{z-\eta}{\beta}\right\}, \quad z>\eta, \beta>0
$$
with parameter estimates as a function of sample statistics
$$
\hat{\alpha}=\frac{4}{s_*^2}, \quad \hat{\beta}=\sqrt{\frac{v_*}{\alpha}}, \quad \hat{\eta}=m_*-\alpha \beta
$$

\end{document}